\documentclass[aps, twocolumn, superscriptaddress, showpacs, nofootinbib]{revtex4-2}

\usepackage[utf8]{inputenc}
\usepackage[T1]{fontenc}
\usepackage{ae,aecompl} 
\usepackage{graphicx}
\usepackage{amsmath}
\usepackage{color}
\usepackage{amssymb}
\usepackage{latexsym}
\usepackage{wasysym}
\usepackage{psfrag}
\usepackage{ifthen}
\usepackage[citecolor=blue,colorlinks=true]{hyperref}
\usepackage{float}
\usepackage[utf8]{inputenc}
\usepackage{lineno}

\usepackage{xcolor}
\usepackage{booktabs}
\usepackage[normalem]{ulem}
\usepackage{placeins}
\usepackage{subcaption}
\usepackage{cleveref}
\usepackage [english]{babel}
\usepackage [autostyle, english = american]{csquotes}
\MakeOuterQuote{"}

\definecolor{bbsalmon}{rgb}{1.0, 0.47, 0.42}
\definecolor{datablue}{rgb}{0.0, 0.0, 1.0}

\begin{document}


\title{Probing eccentric higher-order modes through an effective chirp-mass model}

\author{Ravikiran Hegde}
\email{ravikiran19@iisertvm.ac.in}
\affiliation{School of Physics, Indian Institute of Science Education and Research Thiruvananthapuram, Vithura, Kerala 695551, India}
\author{Nirban Bose}
\email{nirban@iitb.ac.in}
\author{Archana Pai}%
\email{archanap@iitb.ac.in}
\affiliation{
Department of Physics, Indian Institute of Technology Bombay, Mumbai, Maharashtra 400076, India}

\begin{abstract}
The three observational runs of advanced LIGO and Virgo detectors have detected $\sim$ 90 compact binary coalescence events with most of them being quasicircular compact binary merger events. Astrophysical models predict compact binary mergers with appreciable nonzero eccentricity if in dense stellar environments like globular clusters, active galactic nuclei discs. Gravitational wave searches from eccentric systems are either limited to very low eccentricities or rely on model-agnostic time-frequency-based searches. Typically, gravitational waves from eccentric systems with moderate eccentricities carry non-negligible energy in the eccentric higher-order modes. In this work, using the effective chirp mass model [N. Bose and A. Pai, \href{https://doi.org/10.1103/PhysRevD.104.124021}{Phys. Rev. D 104, 124021 (2021)}] developed using EccentricTD waveform, we develop a phenomenological scaling relation for the eccentric higher-order mode time-frequency track. Further, we incorporate this additional information from the higher-order eccentric mode for nonspinning, stellar mass eccentric binaries to improve the constraint on eccentricity. 
\end{abstract}

\maketitle

\section{Introduction}

Gravitational wave(GW) astronomy has been growing rapidly since its first detection in 2015, courtesy of the detection of gravitational wave signals from mergers of compact binary objects like binary black holes, binary neutron stars, and neutron star black hole systems by the advanced ground-based GW detectors like Advanced LIGO {\cite{TheLIGOScientific:2014jea}}, Virgo~\cite{TheVirgo:2014hva}. The detection of nearly $\sim$ 90 compact binary coalescence mergers in the three very successfully conducted observational runs of Advanced LIGO and Advanced Virgo detectors \cite{PhysRevX.9.031040, PhysRevX.11.021053, PhysRevLett.125.101102,GW150914-DETECTION,GW170608-DETECTION,GW170817-DETECTION,LIGOScientific:2020stg,Abbott:2020khf} have thrown light on binary black hole populations, formation channels, and progenitors of short Gamma-ray bursts 
{\cite{10.1088/2514-3433/aae164ch6}}. With the increasing sensitivity of these Advanced GW detectors along with KAGRA \cite{Akutsu:2020his} and LIGO India \cite{Ligo_India,saleem2021science}, we expect to observe many more compact binary merger events in the coming decade \cite{livingrev}. The majority of the detections so far have been quasicircular compact binaries and have shown little to no evidence of eccentricity. There have been constraints placed on the eccentricity of a few detected events \cite{gayathri2022eccentricity, Romero_Shaw_2020,Nitz_2020,Romero_Shaw_2019}. While we have not observed clear evidence of an eccentric compact binary system, with improved sensitivities of the Advanced GW detectors, detection of an eccentric binary is becoming more and more promising. In fact, in \cite{Rodriguez:2017pec}, the study suggests that a significant fraction of the compact binaries retain non-negligible eccentricities at the time they enter the sensitive detection band of the detectors. 

Usually harbingered in dense stellar environments like globular clusters, galactic nuclei, etc, the eccentric binaries carry distinct signatures of formation channels \cite{{Rodriguez:2017pec},{Zevin_2021},{2019ApJ...871...91Z}}. Multiple subchannels can be associated with the dynamic formation channel including, but not limited to, dynamical encounters between black holes, close flybys between binary systems and binary single systems, and perturbation of a binary by a third heavy black hole via Kozai-Lidov mechanism. Eccentric binary black hole (eBBH) systems can also be formed in discs of active galactic nuclei (AGN) due to complex interactions with the gas in the AGN discs \cite{Gayathri_2021}. Each of these subchannels is associated with a characteristic eccentricity distribution \cite{PhysRevD.98.083028}. Further, there have been studies to measure the precession and eccentricity in the in heavy black hole binaries \cite{PhysRevD.107.103049}. In \cite{Zevin_2021}, using state-of-the-art cluster models, it was shown that the increased number of detectable eccentric merger events can significantly constrain the formation channel of the black holes. 

Detection of noncircular and precessing compact binaries is a challenging problem. As far as the model-based matched filter searches are concerned, we are limited by the availability of a suitable template bank for eBBH systems \cite{Nitz_2017}. Although there are efforts to develop template-based searches for eccentric binary neutron star systems for dominant mode GW frequency \cite{Nitz_2020}. Suitable waveforms that carry the complete physics of eccentricity and spins are not ready yet. There are multiple efforts to develop eccentric waveforms. In \cite{PhysRevD.93.064031}, a time domain, inspiral-only waveform, EccentricTD, was developed. This waveform is valid for nonspinning eccentric systems, up to high eccentricities, and includes post-Newtonian (PN) corrections up to 2PN order. 
In \cite{PhysRevD.103.124053}, a surrogate model for eccentric systems encompassing the full inspiral-merger-ringdown phase has been developed. This covers low to moderate eccentricities (up to $\sim 0.25$) and is valid for nonspinning systems and mass ratios up to 5. ENIGMA - another time domain inspiral-merger-ringdown waveform model describing nonspinning binary black hole systems, valid for low to moderate eccentricities, was developed in \cite{PhysRevD.97.024031}. This is valid for low to moderate eccentricities $\sim 0.25$ and for mass ratios $< 5.5$.

An effective one-body formalism has been used to develop time-domain waveforms such as SEOBNRe \cite{SEOBNRe} and TEOBResumS \cite{TEOBResumS} which include the effects of both eccentricity and spin. SEOBNRe is valid for low eccentricities up to $0.2$, and low effective spins up to 0.5. SEOBNRv4EHM; a full inspiral-merger-ringdown time-domain waveform for binary black holes with nonprecessing spins within the effective-one-body formalism, valid for low to moderate eccentricities up to $0.3$ was developed in \cite{PhysRevD.105.044035}. TEOBResumS was generalised to a highly faithful eccentric inspiral model TEOBResumS-DALI \cite{Chiaramello_2020, wf_extend}, both of which are valid for low to moderate eccentricities up to $0.3$. TEOBResumS-DALI is valid for generic orbits, including scattering and dynamically bounded cases(i.e. starting with hyperbolic initial conditions, but merging due to radiation reaction), spinning binaries and contains higher-mode contributions. The inclusion of higher PN terms and alternative modeling strategies to reach improved accuracy were presented in \cite{Placidi_2022, Albanesi_2022_newavenue, carullo2023unveiling}. Parameter estimation of detected events using this model was presented in \cite{iglesias2023eccentricity} with no evidence of any presence of eccentricity. 

Studies have been performed regarding eBBH search in the first and second observational runs using model agnostic coherent WaveBurst algorithm \cite{PhysRevD.93.042004, drago2021coherent, Abbott_2019_ebbh}. Although no eccentric events were found, the study ruled out formation channels with rates $100$ $\mathrm{Gpc}^{-3}\mathrm{Yr}^{-1}$ for $e > 0.1$ corresponding to a black hole mass spectrum with power-law index $2$. There has been waveform-independent analysis \cite{LIGOScientific:2023lpe} to search for eccentric binaries in the third observational run of advanced LIGO Virgo detectors. Although there were no new high-significant candidates apart from those already identified with quasicircular binaries, the study placed an upper limit on the merger rate densities of low to moderate, $(0 \leq e \leq 0.3)$, high-mass binaries (total mass $> 70 \ \text{M}_{\odot}$) to be $0.33\ \mathrm{Gpc}^{-3}\mathrm{Yr}^{-1}$ with $90 \%$ confidence. In a more recent study, the authors have shown that for chirp mass $(>20 \ \text{M}_{\odot})$ and eccentricity $<0.3$, the PyCBC based matched filtered search performs better than the unmodeled coherent WaveBurst search \cite{PhysRevD.102.043005}.

Amongst all the detected events, GW190521 showed some evidence of eccentricity \cite{PhysRevLett.125.101102, Romero-Shaw_2020, gayathri2022eccentricity} which is under discussion and further investigation. In \cite{agn_nature, 2020A&A...638A.119G}, there were efforts to explain the GW190521 event in the light of the AGN disc formation channel. Another study \cite{gamba2022gw190521}, provided evidence for the GW190521 event being formed from the dynamical capture of two stellar-mass nonspinning black holes. An explicit application to the generic case was presented in \cite{Nagar_2021}, while parameter estimation of GW190521, this eccentric inspiral model was introduced. 

In \cite{PhysRevD.104.124021} we took an alternative 
approach. Our aim is to device a tool to localize the features of the eccentricity of the underlying compact binary black hole signal in the time-frequency representation. Using Q-transform, we computed the time-frequency(TF) representation of the GW signal from the nonspinning eccentric binary. It consists of several TF tracks that correspond to the higher-order eccentric modes. We showed that a parameter, similar to the chirp mass, termed as the effective chirp mass $\mathcal{M}_e$, is responsible for the temporal evolution of the lowest (fundamental) TF track. This effective chirp mass parameter reduces to the chirp mass parameter ${\cal{M}}$ in case of low eccentricities or circular binaries. Thus, the lowest TF track, which represents the evolution of the non-eccentric part of the fundamental mode, is termed as the fundamental track. Using this model, we were able to place moderate constraints on the eccentricity of nonspinning, stellar mass symmetric systems for low to moderate eccentricity.

In the case of low eccentricities, the fundamental mode is the most dominant mode. However, for moderate to high eccentricities, a rich structure of higher eccentric modes appears in the time-frequency morphology of the signal. While in the current advanced detectors, the contribution of the higher-order eccentric modes may not be significant, we expect to observe a rich structure of higher-order eccentric modes in gravitational waves emitted by the eccentric binary system as the sensitivity of the advanced ground-based gravitational wave detectors improves. In this work, we extend the previously developed effective chirp mass model as well as the approach to higher-order eccentric modes. We further develop a phenomenological model for the scaling factor of the first two eccentric higher-order modes appeared in the TF representation. We apply the scaling factor to incorporate signal energy from these higher-order eccentric modes and further constraint the eccentricity of the nonspinning, stellar mass eccentric binary system.

The outline of the paper is as follows. In Sec. \ref{Effective Chirp Mass Model (ECMM)}, we give an overview of the Effective Chirp Mass Model (ECMM); in Sec. \ref{Review of the model} we review the model and in Sec. \ref{Stringent validity region for ECMM} we discuss the stringent validity region for the ECMM. In Sec.\ref{Eccentric higher harmonics through ECMM}, we discuss the higher-order eccentric modes whereas in Sec. \ref{TF representation of eccentric higher harmonics} we describe the TF representation of the higher-order eccentric modes. Subsequently, in Secs. \ref{Energy ratio} and \ref{Scaling factor}, we explore the ratio of energy stored in the higher-order eccentric modes for nonspinning, stellar mass eBBH systems and how we use the ECMM to develop a scaling relation for the eBBH systems respectively. In Sec. \ref{Improving constraints on eccentricity using eccentric higher-order modes}, we show how incorporating the higher-order eccentric modes and using the ECMM we improve the eccentricity constraints for the nonspinning, stellar mass eBBH systems. Finally, we discuss the implications of our work and lay down the conclusions and future plans in Sec. \ref{conclusions}.

\section{Effective Chirp Mass Model}
\label{Effective Chirp Mass Model (ECMM)}
 In this section, we provide the overview of the ECMM developed in \cite{PhysRevD.104.124021}.
\subsection{Salient features the model}
\label{Review of the model}

The ECMM \cite{PhysRevD.104.124021} is based on Peters and Mathews \cite{peters} slow motion and adiabatic 
approximation. For nonspinning, stellar mass eccentric-BBH systems -- the equations governing the rate of change of instantaneous eccentricity, frequency and the size of the semimajor axis, when evolved numerically along with the rate of change of energy, we obtained the inspiral frequency evolution as
\begin{equation}
    \pi^{8/3} \frac{96}{5} {\left(\frac{G\mathcal{M}_e}{c^3}\right)}^{5/3} (t_c-t) - \frac{3}{8} f(2,2)(t)^{-8/3}=0 \ ,
  \label{chirpmassE}
\end{equation}
where $G,c$ are the universal gravitational constant and velocity of light respectively, $t_c$ is the time of the coalescence of the binary, and $\mathcal{M}_e$ is called the effective chirp mass parameter. This effective chirp mass for eBBH systems captures the frequency evolution of the
$(2, 2)$ mode with effects of eccentricity at the Newtonian order. It does not account for any additional general relativity effects. The corresponding phenomenological model $\mathcal{M}_e$ is in terms of chirp mass $\mathcal{M}$ and eccentricity $e$ of the system at a fiducial frequency of 10 Hz and is given by 

\begin{eqnarray}
 {\mathcal{M}_{e}} &=& {\mathcal{M}} \bigg( 1 + \alpha(\mathcal{M}) {e}^2 + \beta(\mathcal{M}) {e}^4 + \gamma(\mathcal{M}) {e}^6 \bigg) \nonumber , \\
 &\equiv & {\mathcal{M}} {\mathcal G}(e)\ .
 \label{phenomenological model eqn}
\end{eqnarray}
The parameters $\alpha(\mathcal{M}),\beta(\mathcal{M}),\gamma(\mathcal{M})$ are polynomials in $\mathcal{M}$.
For low eccentricities, the effective chirp mass reduces to the chirp mass of the system. As the eccentricity increases, the system radiates energy at a higher rate due to asymmetry associated with the eccentricity thereby taking less time to coalesce, thus increasing the value of ${\mathcal{M}_{e}}$.

\begin{figure*}[htb!]
 \centering
 \begin{subfigure}[b]{0.55\linewidth}
  \includegraphics[width=\linewidth]{ 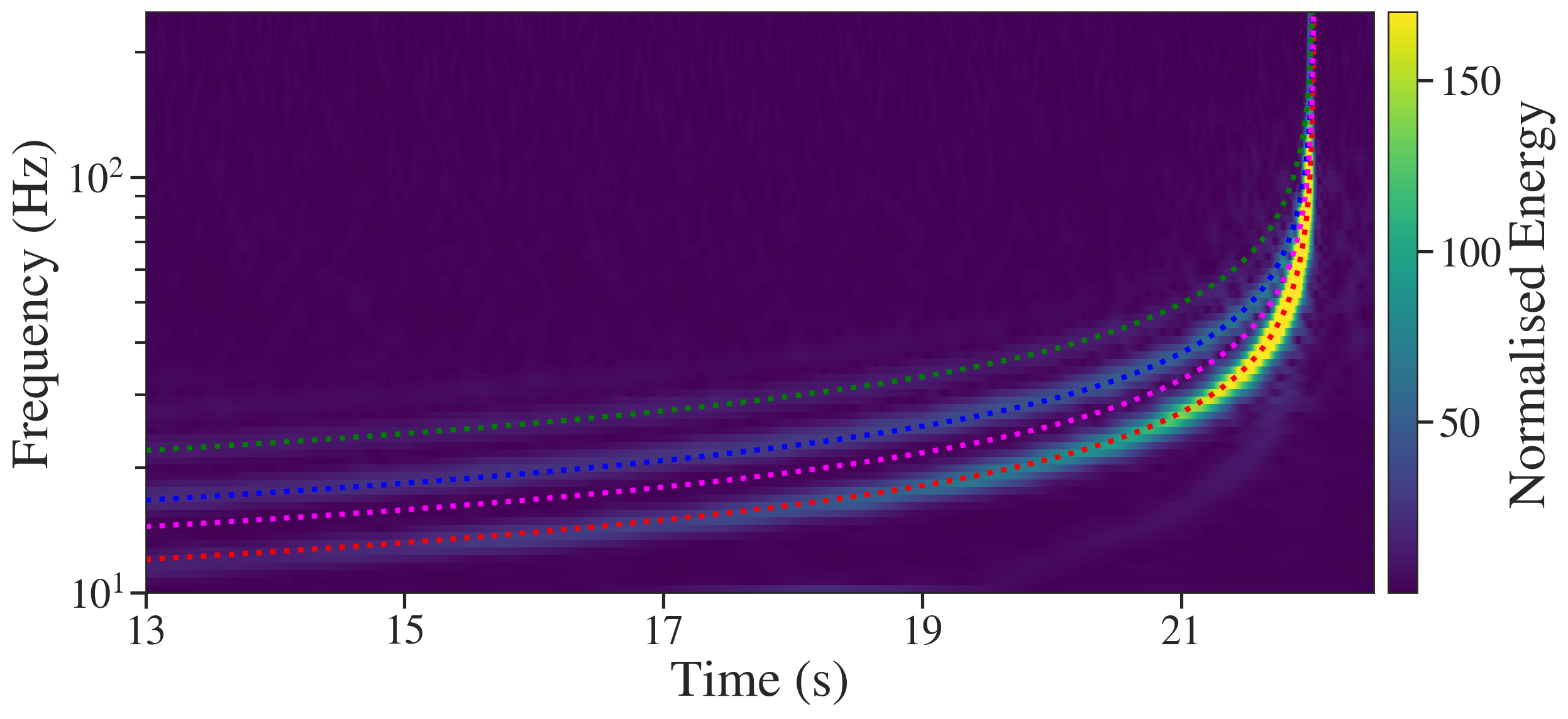}
  \caption{TF map from the Q-transform.}
  \label{subfig:tf_map}
 \end{subfigure}
 \hspace{0.5cm}
 \begin{subfigure}[b]{0.35\linewidth}
  \includegraphics[width=\linewidth]{ 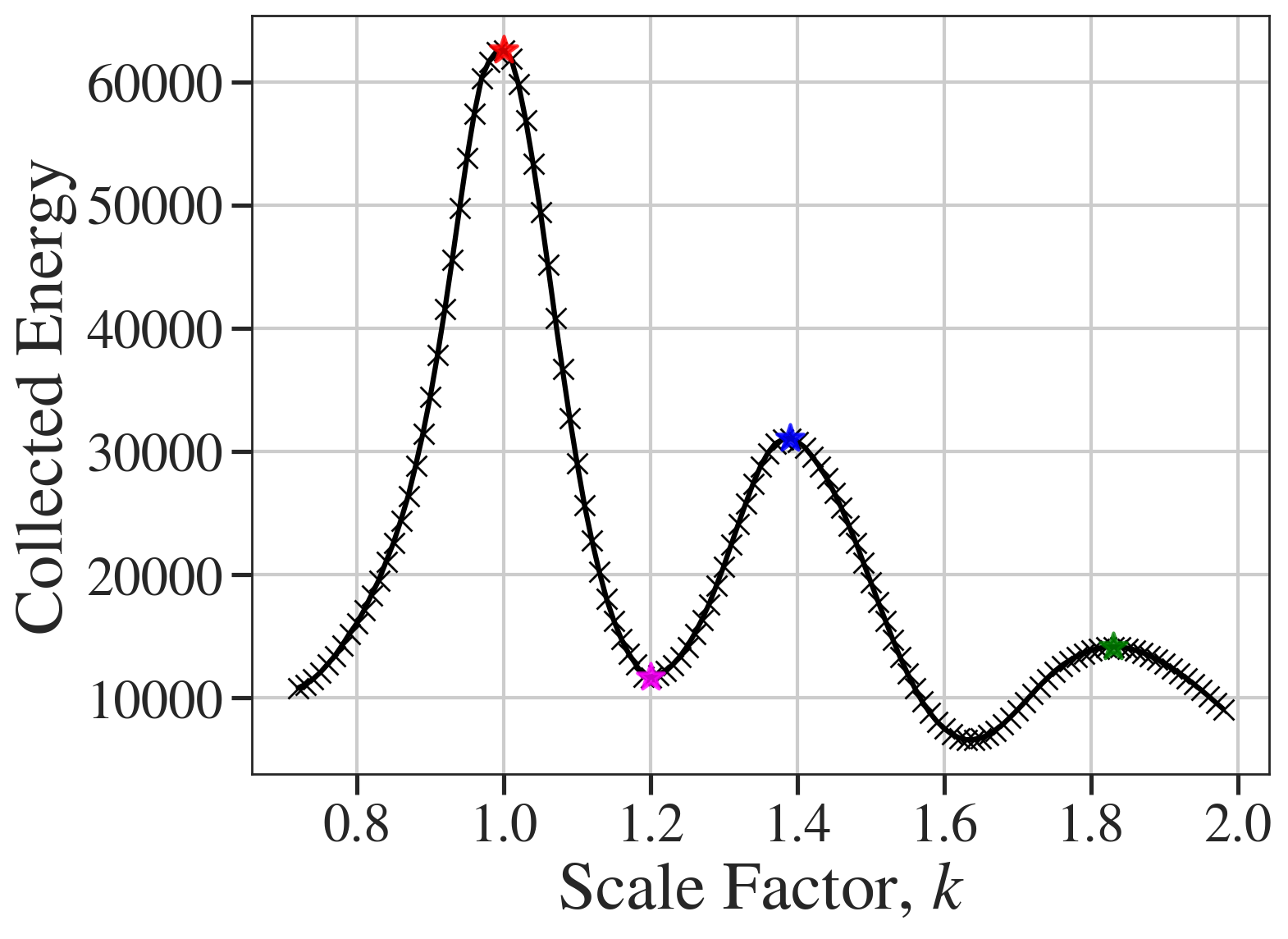}
  \caption{Collected energy vs scale factor $k$.}
  \label{subfig:energy_vs_scale}
 \end{subfigure}
 \caption{Illustration with a $(\mathcal{M},e)\sim (15 \ \text{M}_{\odot},0.2)$ system. The lowest track in the TF representation corresponds to the fundamental track. The other higher tracks correspond to higher-order eccentric modes. The red dotted lines represent frequency evolution for the fundamental mode and the blue and green dashed lines represent the first and second eccentric higher-order mode obtained from the ECMM and the scaling factors corresponding to the peaks in collected energy. The pink dashed line does not correspond to any eccentric mode of the system, as can also be seen by the low collected energy. }
 \label{fig:TF_map_energy_vs_scale}
\end{figure*}

We compared this model of the frequency evolution with the time-frequency representation of the EccentricTD waveform. We then match the morphology of the fundamental mode with the ECMM in the time-frequency representation. We showed in \cite{PhysRevD.104.124021} that the frequency evolution computed above matches with the lowest TF track of the GW signal emitted by the nonspinning eccentric binary system modeled with the EccentricTD waveform model. As an example, Figure \ref{subfig:tf_map} shows the TF representation obtained by the Q-transform of the EccentricTF signal. The lowest $(2, 2)$ mode TF track overlaps with the frequency evolution shown by the ECMM as the dotted black line. Henceforth, we denote this fundamental track as $f_o(2,2)$.

\subsection{Validity region of ECMM for higher-order eccentricity modes}
\label{Stringent validity region for ECMM}

As mentioned earlier, we use the Q-transform to obtain the TF representation of the time domain waveform \cite{shouravthesis}. The Q-transform is similar to the wavelet transform and it projects the signal into a multiresolution time-frequency representation. It is characterized by central time, central frequency, and Q parameter. For low values of Q, the time resolution of pixels is smaller while the frequency resolution is larger. Due to the finite size of the TF pixel, we expect deviation between the TF path given by the ECMM and the frequency track extracted from the TF pixels from the Q-transform of the EccentricTD waveform. This might be tolerable if we are considering only the dominant fundamental mode. However, for higher-order eccentric modes, we need to be more careful while collecting the TF pixels corresponding to the chosen TF track with our model due to the proximity of the tracks. Hence we demand stringent conditions on the validity region of the ECMM.

We consider eBBH systems with chirp mass and eccentricity uniformly sampled in the range 5$\ \text{M}_{\odot}$ -- 35$\ \text{M}_{\odot}$ and 0 -- 0.4 respectively. Using these systems we simulate eBBH systems with EccentricTD waveform and inject in Gaussian noise whitened with Aplus power spectral density \cite{Maggiore_2020}. We project the time series signal in the TF representation using Q-transform. We then consider the frequency evolution of the eBBH systems using our effective chirp mass model from  \cite{PhysRevD.104.124021}. 
For each of the selected pixels from the dominant, fundamental TF track, we compute the mean square deviation on the frequency deviation metric 
\begin{equation}
\Delta f = \sqrt{\frac{\Sigma_{i=1}^{P} {(f_{(\mathrm{Qtransform})_{i}} - f_{(\mathrm{model})_{i}})^2}}{N}}\ . 
\end{equation}
Here, $f_{({\mathrm{Qtransform}})}$ denotes the frequency of the extracted pixel corresponding to the fundamental track from the Q-transform and $f_{(\mathrm{model})}$ denotes the frequency predicted by the model, $i$ denotes the time bin and $N$ denotes the total number of extracted pixels. We choose $\Delta f = 3\ \mathrm{Hz}$.

We observe that for low-mass and low-eccentricity systems, the deviation metric shows a much lower value, whereas for high-mass and high-eccentricity systems, the deviation metric increases indicating lower efficacy of the effective chirp mass model for those systems. In those cases, the model deviates towards the late inspiral and near merger regimes thereby contributing towards larger values to the deviation metric. Putting the stringent cut of 3 Hz on the deviation metric, a curve representing those binary systems that satisfy this condition is

\begin{equation}
  \mathcal{M}=\sqrt{94.02/e -1}.
  \label{validity eqn}
\end{equation}
If this cut is relaxed to 5 Hz, the model holds for the original regime of chirp mass $5 \ \text{M}_{\odot}$ -- $35 \ \text{M}_{\odot}$ and eccentricity up to $0.6$ \cite{PhysRevD.104.124021}.

\section{Eccentric higher modes and ECMM}
\label{Eccentric higher harmonics through ECMM}
In this section, we compute the TF domain representation of the dominant quadrupolar, gravitational wave signals from the eccentric, symmetric and nonspinning binary black hole systems. We explore the higher-order eccentric modes in the TF domain using ECMM.

\subsection{Eccentric higher modes for noncircular binary}
\label{TF representation of eccentric higher harmonics}

For Keplerian motion in a binary system having an elliptic orbit, the trajectory of the reduced mass in the $x$--$y$ plane is a periodic function that can be decomposed via Fourier decomposition. Therefore, when we consider the adiabatic evolution of the binary orbit and consequently the gravitational waves emitted from such noncircular orbits, the orbital information gets naturally translated into the respective higher-order eccentric modes. 

We use EccentricTD waveform \cite{PhysRevD.93.064031}, to simulate the eccentric higher-order mode signals. EccentricTD waveform is a time-domain, inspiral-only approximant that models eccentric binaries with nonspinning components. It is an eccentric extension of the quasicircular TaylorT4 \cite{PhysRevD.80.084043} approximant up to 2PN order. It considers a 2PN-accurate emission of gravitational waves from compact binaries inspiraling along 2PN-accurate eccentric orbits described by a Keplerian-type parametrization. Hence, including the effect of orbital motion, periastron precession, and radiation reaction on the emitted gravitational waves. Moreover, an "exact-in-eccentric" approach to describe the binary dynamics makes it suitable for modeling systems having initial eccentricities up to 0.9.

Typically, for low to moderate eccentricities, the fundamental $f_o(2,2)$ mode is the most dominant mode. Often for very low eccentricity, it corresponds to the circular case.
In such cases, the higher eccentric modes do not have much energy contribution. As the eccentricity increases, the contribution from higher-order eccentric modes starts to contribute more. 

Figure \ref{subfig:tf_map} shows the TF representation of an eccentric binary system with $(\mathcal{M}, e) \sim 15 \ \text{M}_{\odot},\ 0.2$ using the Q-transform. The presence of multiple TF tracks is clearly visible corresponding to the higher-order eccentric modes. The lowest TF track corresponds to the fundamental track $f_o(2,2)$ whereas the next visible track corresponds to the first higher-order eccentric mode. The dotted black line represents the fit obtained from the ECMM which we discuss in Sec. \ref{Scaling factor}. We expect that as the sensitivity of the detectors improves, the higher-order eccentric modes will become more and more {\it evident} in the TF representation.

\subsection{Scaling relation}
\label{Scaling factor}

\begin{figure*}[htb!]
 \centering
 \begin{subfigure}[b]{0.48\linewidth}
  \includegraphics[width=\linewidth]{ 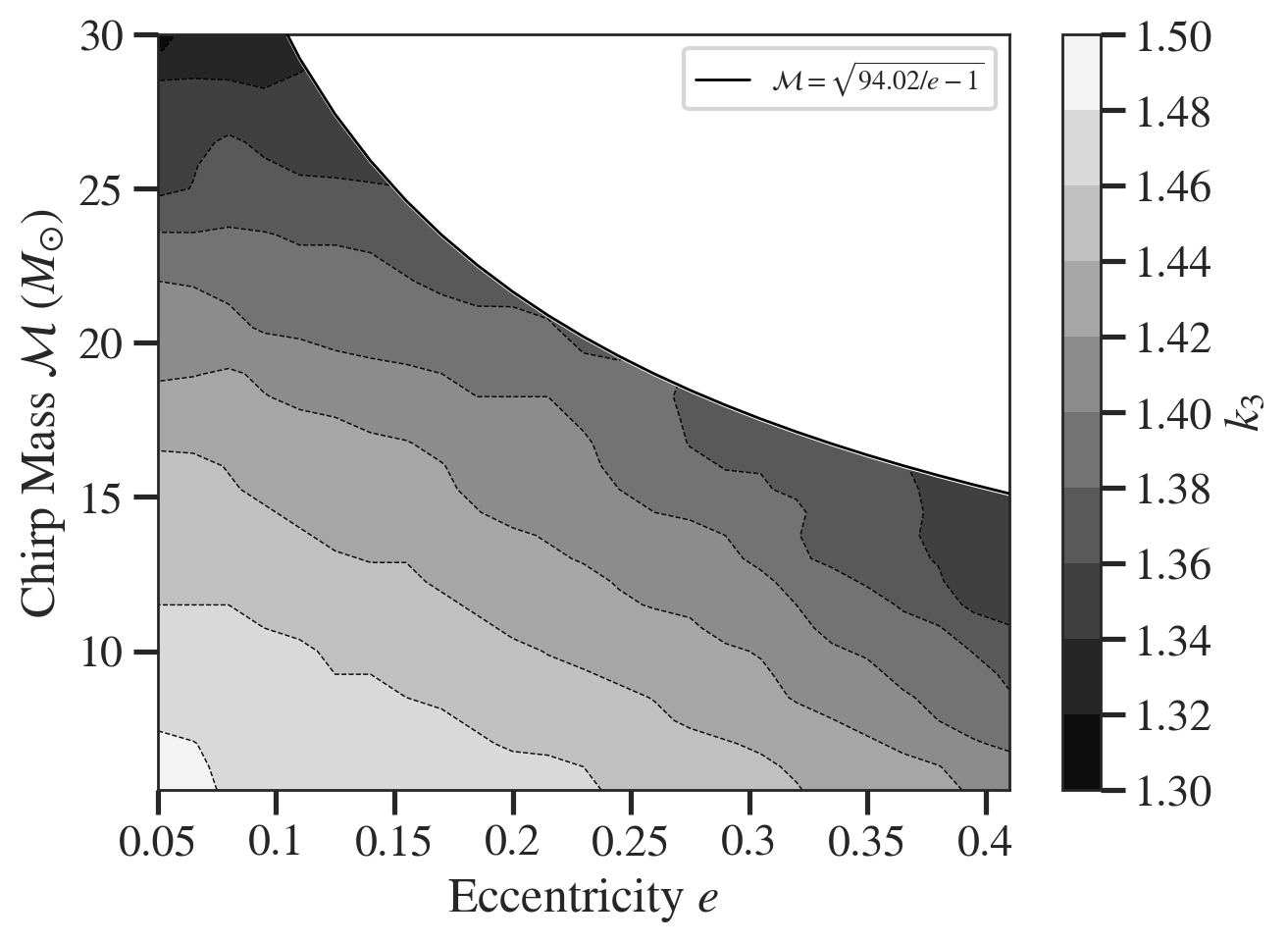}
  \caption{The scale factor $k_3$}
  \label{subfig:scaling_relation}
 \end{subfigure}
 \hspace{0.5cm}
 \begin{subfigure}[b]{0.48\linewidth}
  \includegraphics[width=\linewidth]{ 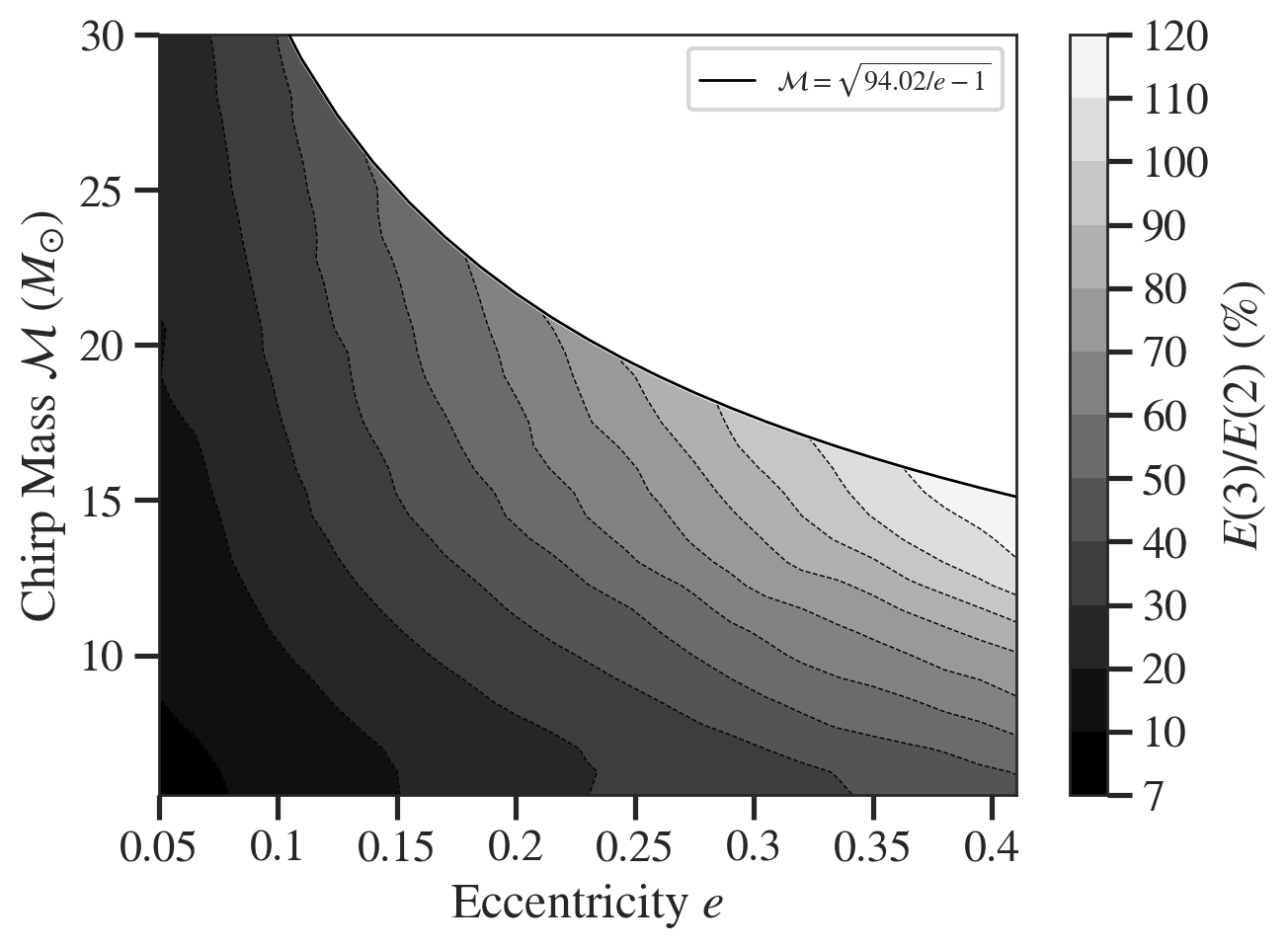}
  \caption{Energy ratio for $n=3$ track}
  \label{subfig:energy_ratio_plot}
 \end{subfigure}
   \caption{Contours for obtained values for the systems in the validity region of ECMM(depicted by the solid black line).The energy ratio is the ratio of energy along the first higher-order eccentric track $E(3)$, and the energy along fundamental track $E(2)$. }
 \label{fig:energy_and_scaling}
\end{figure*}

In this section, we study and model the first two higher-order eccentric tracks for the case of eccentric nonspinning binary systems. Typically, we expect that the gravitational wave frequency of the higher-order eccentric track is related to the fundamental track by a scaling relation, $f(n>2) \equiv k_n f_o(2,2)$, where $k_n$ is the scaling factor that depends on $n$, and $f_o(2,2)$ is the time-varying frequency of the fundamental track.  For example, gravitational wave signals from nonspinning, circular binaries with asymmetric mass ratios emit higher-order modes. They often arise due to asymmetry in the masses. Typically, $(3,3)$ mode is the next dominant higher-order mode, and the frequency of this mode is $3/2$ times the $f(2,2)$. Of course, in that case, the higher modes arise from the expansion of the signal in the spin-weighted spherical harmonics. In the case of eccentricity, the modes arise from the orbital shape and the general relativity effects such as a periastron advance in the waveform approximation. Therefore, it is not obvious if the frequency of the next-order eccentricity mode would be $3/2$ times that of $f_o(2,2)$. As the rate of periastron advance is governed by the chirp mass and eccentricity of the system \cite{Tiwari_2019}, we expect the scaling relation to be dependent on the system; namely the eccentricity parameter and the chirp mass. 

In order to obtain the phenomenological model for the scaling relation, we compute the Q-transform of the time series data of the injected EccentricTD gravitational wave signals of a given chirp mass and eccentricity with equal mass systems. We have the ECMM for the fundamental track $f_o(2,2)(t)$ following Eq. \ref{chirpmassE}. 

To probe the scaling factor, we compute the analytical tracks as $k f_o(2,2)(t)$, where $k$ ranges from 0 to 2, with an increment of 0.01. For each analytical track, the total energy along the track is calculated from the TF pixels underlying the track. 
The mean energy of the central pixel (closest to $f_o(2,2)$ at the time instance) and neighboring pixels corresponding to the frequency of the analytical track at each time instance is summed to get the total energy along the track. The number of neighboring pixels chosen depends on the width of the track in the Q-transform. To include the bright pixels from the fundamental track, we include five pixels (two on each side of the central pixel). For the first eccentric higher-order track, which has a narrower width, we use three pixels (one on each side of the central pixel). We limit the addition of energy in the first eccentric higher-order track to pixels corresponding to frequencies up to approximately $0.8\times F_{LSO}$ (frequency of last stable orbit for the circular case). This is to avoid double counting of the pixels, as the fundamental track and the first eccentric track merge further closer to coalescence.

Figure \ref{subfig:energy_vs_scale} shows the collected energy along the track as a function of the scale factor for the ECMM. We note that the first peak is at $k=1$ which corresponds to the fundamental $f_o(2,2)$ mode. As we increase the value of $k$, it can be seen that the second peak corresponds to that value of $k$ when the analytical track matches with the first higher-order eccentric TF track. In Figure \ref{subfig:tf_map}, the first higher-order eccentric track is captured by $k=1.39$; corresponding to the second peak in Figure \ref{subfig:energy_vs_scale}. The value of $k \neq 3/2$ supports the earlier argument.

We consider $\sim$680 systems uniformly sampled in chirp masses in ($5 \ \text{M}_{\odot}$ -- $35 \ \text{M}_{\odot}$) and eccentricity in (0.01 --0.4) from the region of validity of the ECMM according to Eq. \ref{validity eqn}. For each system, we simulate eBBH signals using EccentricTD waveform model\cite{PhysRevD.93.064031} and inject them in simulated Gaussian noise colored by Aplus noise \cite{Maggiore_2020}. The optimum signal-to-noise ratio(SNR) of these systems was fixed at a high value of 1000 to ensure redundancy of any systematic effects and sufficient power in the higher eccentric mode. Please note this choice of SNR was just to build the accurate phenomenological model for the scale factor $k_3$ 
and has no observational biases. The scaling factor corresponding to the first eccentric higher-order TF track is computed for these systems as explained earlier in this section.

Figure \ref{subfig:scaling_relation} shows the scaling factor contours for different chirp mass and eccentricity values within the validity region. The color bars represent the scale factor value. For lower chirp mass and eccentricity the scale factor $k_3$ value is close to $3/2$, whereas the scale factor value decreases as we move away towards the higher chirp mass and higher eccentricity values. Thus, it can be seen that the first higher-order eccentric mode does not follow the simple $f(n>2) = \frac{n}{2} f_o(2,2)$, rather the scale factor depends on the chirp mass and eccentricity of the system. The physical reason behind the nonintegral scaling factor is due to the effect of periastron advance in eccentric orbits as predicted in \cite{Tiwari_2019}. In \cite{Tiwari_2019}, it is shown that for eccentric binaries with periastron advance, the higher-order modes get shifted by a certain amount dictated by the rate of periastron advance given. As the rate of periastron advance is governed by the chirp mass and orbital eccentricity, the scaling factor essentially becomes a function of the same.

\begin{figure*}[t]
 \centering
 \begin{subfigure}[b]{0.48\linewidth}
  \includegraphics[width=\linewidth]{ 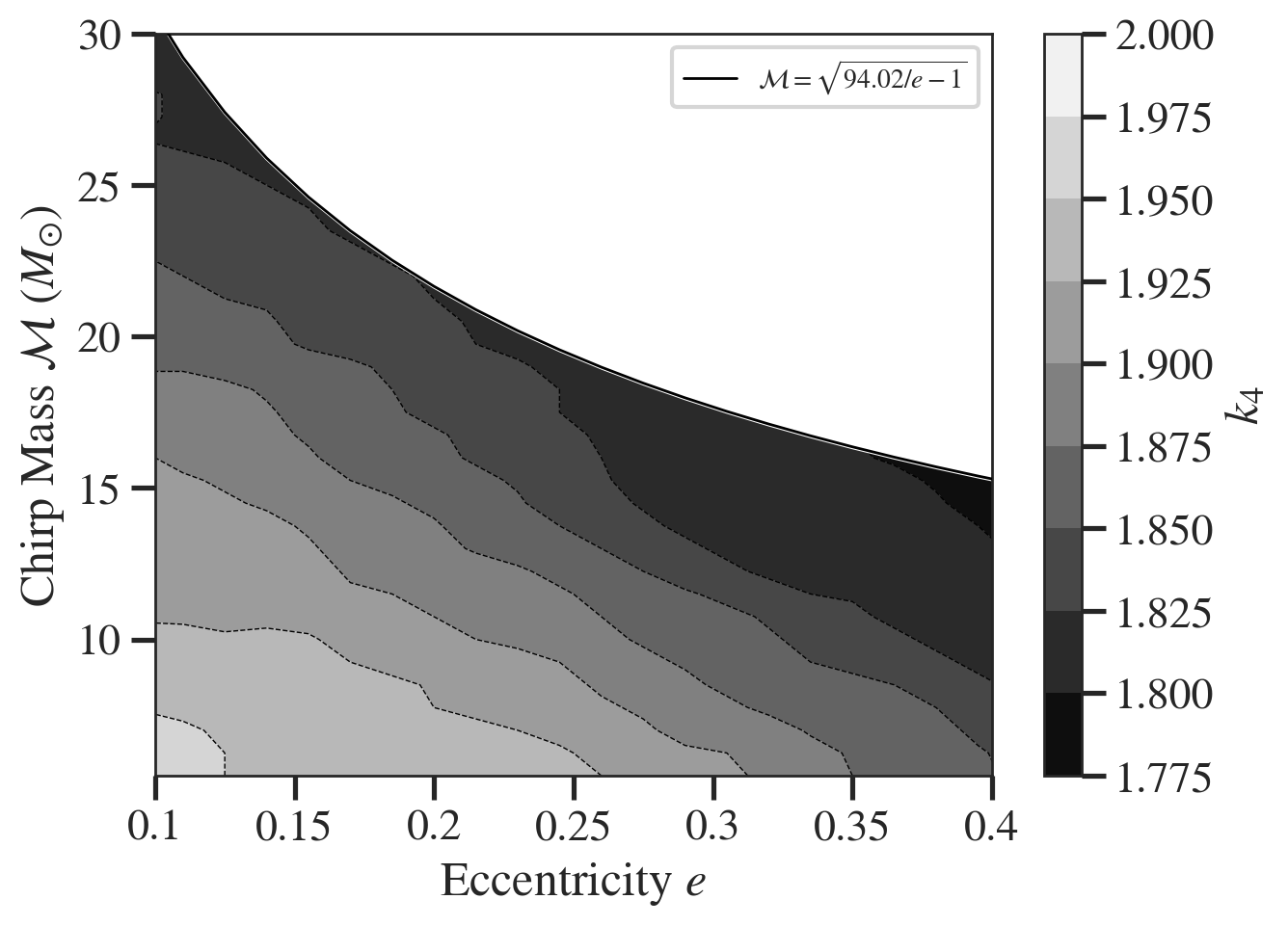}
  \caption{The scale factor $k_4$}
  \label{subfig:scaling_relation_n=4}
 \end{subfigure}
 \hspace{0.5cm}
 \begin{subfigure}[b]{0.48\linewidth}
  \includegraphics[width=\linewidth]{ 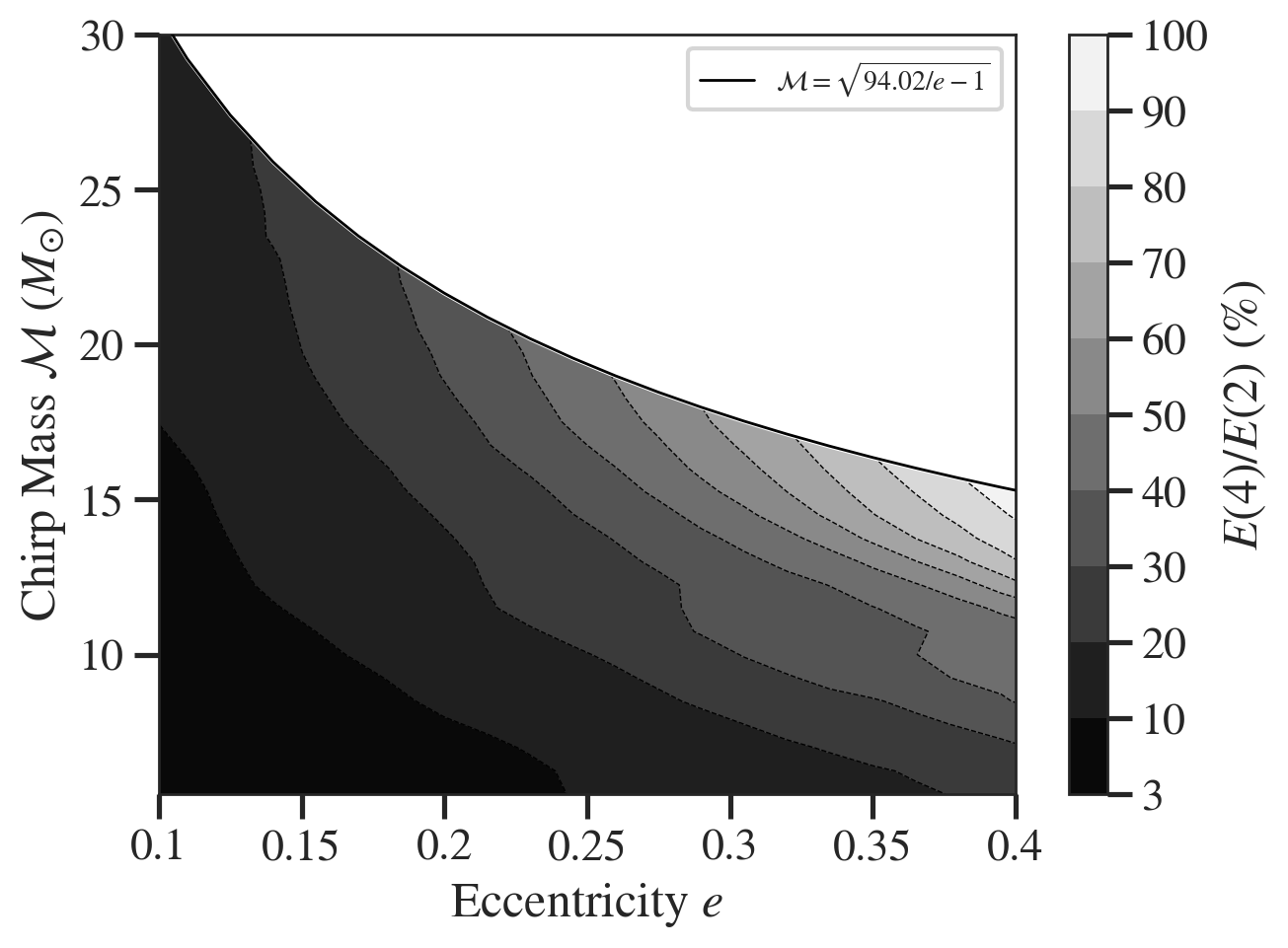}
  \caption{Energy ratio for $n=4$ track}
  \label{subfig:energy_ratio_plot_n=4}
 \end{subfigure}
 \caption{Contours for obtained values for the systems in the validity region of ECMM(depicted by the solid black line).}
 \label{fig:energy_and_scaling_n=4}
\end{figure*}

For high SNR, next-order eccentric tracks become discernible in the time-frequency (TF) representation, as showcased in Figure \ref{subfig:tf_map}. Building on the track-extraction detailed in Sec. \ref{Scaling factor}, we obtain the scaling relation for the next eccentric higher-order mode ($n = 4$). In this computation, we include pixels up to $0.8 \times F_{LSO}$, with one pixel on either side of the central pixel. The calculated scaling factors $k_4$, within the ECMM's validity region, are illustrated in Figure \ref{subfig:scaling_relation_n=4}. When eccentricity and chirp mass are low, $k_4$ approaches 2. Yet, as chirp mass and eccentricity increase, $k_4$ decreases, exhibiting a pattern akin to $k_3$. This behavior aligns with the previously discussed trends.

\subsection{Energy ratio}
\label{Energy ratio}
In Section \ref{Scaling factor}, we obtained the numerical values for $k_3$. As energy in each higher-order eccentric mode is not readily available so far from the exact-in-eccentric implementation of the EccentricTD approximant, we calculate the energy along the track from the fundamental mode using the ECMM and the first higher-order eccentric mode using the scaling relation. We define energy ratio as the ratio of total energy along the higher-order eccentric track and the fundamental track, i.e., $E(n)/E(2)$. 

Figure \ref{subfig:energy_ratio_plot} shows the obtained energy ratio $E(3)/E(2)$ in $(\mathcal{M},\ e)$ space. We can observe that in systems featuring low eccentricity and chirp mass, the energy contained in the first higher-order eccentric mode is proportionally smaller compared to the energy associated with the fundamental mode. However, as the chirp mass or eccentricity increases, the energy ratio increases. For low chirp mass and eccentricities, the first higher-order eccentric mode contains energy as low as $7 \%$ of the fundamental mode. The energy ratio increases significantly to $75 \%$ and above for systems with higher chirp mass and eccentricity. For instance, in a system with $(\mathcal{M},e)= (8\ \text{M}_{\odot}, 0.2)$ the energy ratio is $\sim 35 \%$, whereas for a $(14\ \text{M}_{\odot},\ 0.4)$ system, the energy ratio is as high as $\sim 80 \%$. Figure \ref{subfig:energy_ratio_plot_n=4} plots the energy ratio $E(4)/E(2)$. Similar to $n=3$, $E(4)$ is comparable to $E(2)$ for higher eccentricities. The increasing contribution of higher-order eccentric modes is consistent with the observations in other studies \cite{Tessmer2006}.

\section{Implications of the scaling relation for higher-order eccentric modes}

In this section, we discuss the direct implication of the proposed scaling relation for the eccentric higher-order mode. 
\subsection{Improving constraints on eccentricity }
\label{Improving constraints on eccentricity using eccentric higher-order modes}

In the previous study \cite{PhysRevD.104.124021}, we used the ECMM to place constraints on the eccentricity of a nonspinning, stellar mass eBBH system in the Aplus regime. The study was based on the assumption that the frequency evolution of the fundamental track gets modified owing to the presence of eccentricity. The frequency evolution was modeled using the ECMM, and it was used to impose constraints on the eccentricity. 
Here, we extend our approach to include the higher-order eccentric mode along with the fundamental mode to constrain the eccentricity. In this section, we show that using the scaling factor, we incorporate additional energy from the higher-order eccentric mode and thus improving the constraint on eccentricity.

Given a confident, loud detection, we can first estimate the effective chirp mass from the TF representation. As before, with the estimated chirp mass, we identify the initial patch of $(\mathcal{M},e)$ from the ECMM. Within that patch, we lay down a sufficiently fine grid in the $\mathcal{M}$-$e$ space. Further, for each grid point we identify the scale factor from the scaling relation (Figure \ref{subfig:scaling_relation}) obtained in Sec. \ref{Scaling factor}. The frequency evolution of the fundamental track and the first higher-order eccentric track is then simulated using $k_3$. The TF pixels along the neighborhood of these analytically simulated tracks are collected and the mean energy of the collected pixels is computed as,

\begin{equation}
  E \sim \sum_{i} \|X_i\|^2/N,
\end{equation}
where $i$ denotes the $i$th pixel and $N$ denotes the total number of pixels.
The points in the $\mathcal{M}$-$e$ grid close to the true parameter values are expected to recover higher energy as compared to the other grid points. The presence of any inherent degeneracy due to the functional dependence of $\mathcal{M}_e$ on $\mathcal{M},\ e$ is now reduced due to the incorporation of the first higher-order eccentric mode. Incorporating the higher-order eccentric track adds more information in terms of additional energy and thus reducing the error in the estimated parameters. This naturally constrains the eccentricity better. 
\begin{figure}
\centering
\includegraphics[width = \linewidth]{ 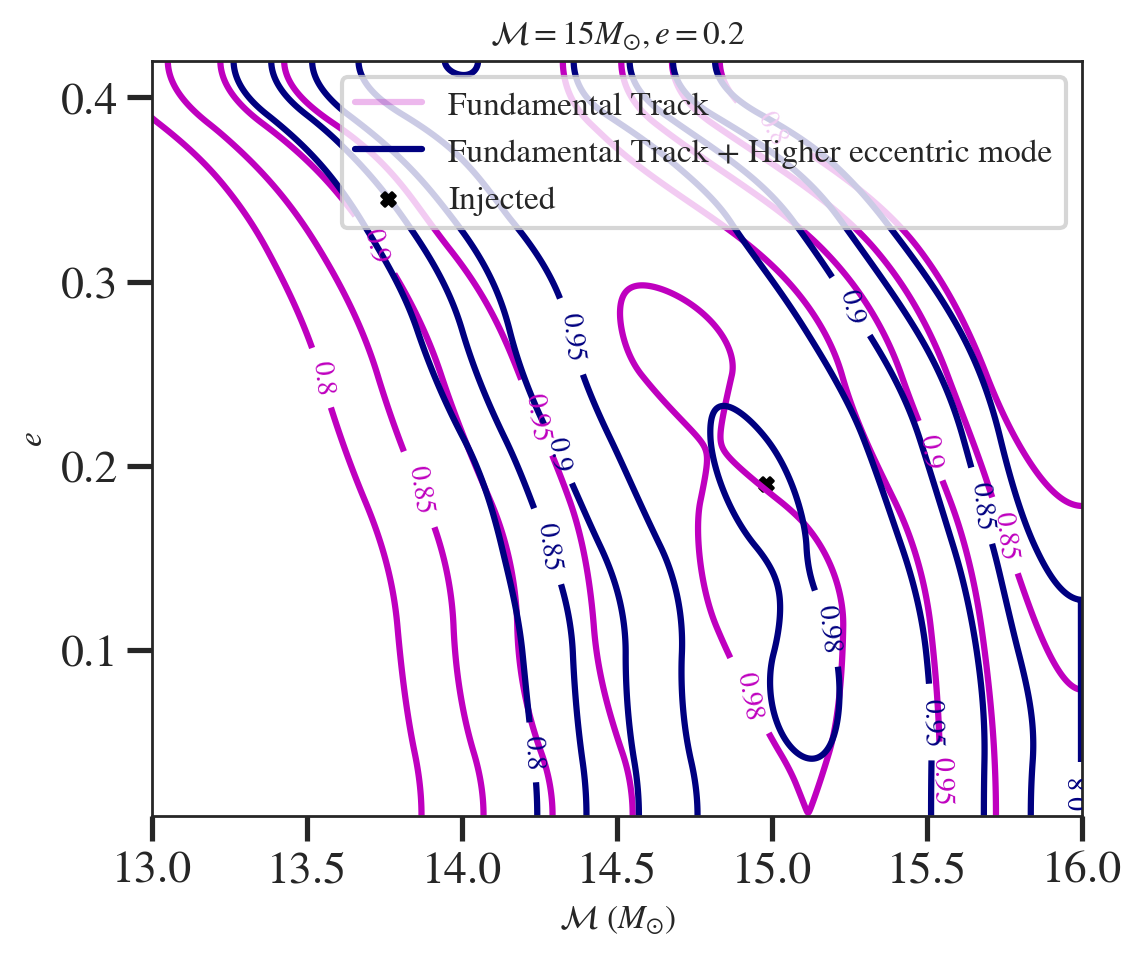}
   \caption{Normalized energy contours of an eccentric system with $(\mathcal{M},e)= (15 \ \text{M}_{\odot},0.2)$. The magenta contours correspond to the case where the energy contribution is from only the fundamental track while the blue contours denote the case where energy contribution comes from both the fundamental and the first higher-order eccentric mode for the same system. The different lines correspond to different normalized energy levels.}
  \label{fundamental_and_eccentric}
\end{figure}
To validate the method, we consider an eBBH system with (${\mathcal{M},\ e}$) as $(15 \ \text{M}_{\odot},\ 0.2)$ respectively. The relative energy stored in the first higher-order eccentric mode is around $\sim 50 \%$ with respect to the fundamental track. We generate GW signal using EccentricTD waveform and inject in simulated Gaussian noise colored by Aplus amplitude spectral density. The optimal SNR of the system is set at $\sim$ 100. Fig \ref{fundamental_and_eccentric}, shows the mean energy contours of the system. The magenta contours correspond to the case where the energy contribution is from only the fundamental track while the blue contours denote the case where energy contribution comes from both the fundamental track as well as the next higher-order eccentric mode for the same system. The different lines correspond to different normalized energy levels. Incorporating the first higher-order eccentric mode along with the fundamental mode clearly shows an improvement of the constraints in the $\mathcal{M},e$ space with respect to considering the fundamental mode only.

To consider the effects of chirp mass and eccentricity on the improvement of constraint, we studied two more systems with $({\mathcal{M}, e})$ as $(18 \ \text{M}_{\odot}, 0.1)$ and $(15 \ \text{M}_{\odot}, 0.35)$ respectively. The relative power stored in the higher-order mode eccentric track with respect to the fundamental track is $\sim 20 \%$ and $\sim 80 \%$, respectively, and thus significantly improves the eccentricity constrain if contribution from higher-order eccentric modes is added. Quantitatively, in terms of mean energy, we have an improvement of nearly $9.74\%$ after incorporating the second track for the $(18 \ \text{M}_{\odot},\ 0.1)$ system, whereas for the $(15 \ \text{M}_{\odot}, \ 0.2)$ system, it is around $22.39\%$. It takes a value of $57.01 \%$ for the $(15 \ \text{M}_{\odot},\ 0.35)$ system. 
 
\subsection{Removing the degeneracy between the higher-order harmonics and higher-order eccentric modes}
The gravitational waves are predominantly emitted from the quasicircular coalescing compact binaries at twice the orbital frequency. However, in general, the emitted gravitational wave signal can be decomposed into the sum of spin-weighted spherical harmonics. Hence apart from $(l,m)=(2,2)$ mode, other higher-order harmonics like $(2,1),(3,2),(3,3),(4,4)$ etc., also carry non-negligible power in them. The effect of these higher-order harmonics becomes more important as the mass asymmetry between the components of a binary system increases. These higher-order harmonics leave their signatures in the TF representation of the signal in terms of TF tracks that correspond to $(3,3)$, $(4,4)$ modes, etc. The scaling relation for such tracks is $f(n>2) \equiv k_n f(2,2)$ where $k_n = n/2$. In fact, searching for this signature in the TF representation of the signal, in \cite{PhysRevD.103.064012}, authors have developed an independent approach to find the evidence of the higher harmonics. Using this method, when applied to GW190412 showed evidence for the higher mode harmonics \cite{LIGOScientific:2020stg}. In another study, \cite{Vedovato_2022} carried out using a minimally modeled search of higher-multipole gravitational-wave radiation to detect the $(3,3)$ multipole in the inspiral phase of GW190814 and GW190412. 

While here we consider symmetric, eccentric systems, the above also holds true for the nonquasicircular systems with asymmetry in masses. We expect that for eccentric, asymmetric systems, higher-order modes both due to asymmetry and eccentricity will show signatures in the signal. If these signatures are same/similar, there would be degeneracy. In this work, we show that for moderate (>0.1) eccentricity, the two scaling factors would differ. Thus, the degeneracy can be broken in such cases. More particularly, our scaling model, shows clearly that $k_n$ for higher-order eccentricity is not a ratio of two integers and also depends on the eccentricity of the system.
One additional implication of our scaling relation is to construct a quick tool to break this degeneracy. Using the difference in scaling factors, one can infer whether there is a non-negligible eccentricity present in the system or whether there is a mass asymmetry associated with the system, or the presence of both. 

\subsection{Implication on spinning systems}
So far, we have developed the scaling relation for the nonspinning, eccentric systems. The main reason behind this is the following. Since our current approach is nonstandard, we need to clearly understand the dependence of the ECMM model and the scaling factor on the two distinct physical parameters such as eccentricity and spins.

In the case of the spinning eccentric system, we test the ECMM and scaling factor model $k_3$, $k_4$ with respect to two waveforms; namely TEOBResumS and SEOBNRe for low spin and eccentricity values. Our tests show that the ECMM model works well with low spins (up to 0.15) and low eccentricity (up to 0.2). Thus, the methodology which we have developed can be used for these systems. 

In order to incorporate high spin values in to the phenomenological model, we need to see the effect of the effective spin $\chi_{eff}$ parameter on the ECMM. In that case, the phenomenological model can be expressed in terms of three parameters; namely chirp mass, eccentricity and the effective spin. We plan to carry this out in future work.

\section{Conclusions and future work}
\label{conclusions}

In \cite{PhysRevD.104.124021}, it was shown that the inspiral frequency evolution as appears in the fundamental track in the TF representation of the quadrupolar GW emitted from the nonspinning, eccentric, symmetric binaries can be represented by the effective chirp mass parameter. This ECMM gives a phenomenological model in terms of the chirp mass and eccentricity of the system measured at the fiducial frequency. Eccentric nonspinning and symmetric mass binaries emit higher-order eccentric modes that as clearly visible in the TF representation of EccentricTD waveform.

In this work, we extend the ECMM to capture the frequency evolution of the higher-order eccentric tracks in the TF representation. Using EccentricTD as an example, we obtain the phenomenological model for the scaling factor which relates the frequency of the first eccentric higher mode track and the lowest fundamental track in the TF morphology. We note that unlike the scaling relation observed between various higher-order harmonics, the scaling factor of eccentric higher-order modes is not a constant (equal to $n/2$) but depends on the chirp mass and eccentricity of the system due to the periastron advance in eccentric orbits. This is the main result of this work.

Using the scaling relation, we incorporate the contribution of the first higher-order eccentric track in obtaining the eccentricity in a heuristic way. We show with an example that, accounting energy from higher-order eccentric mode naturally helps in improving the constraint on the eccentricity as compared to the fundamental track. This clearly shows the ability to improve eccentricity information when we incorporate more and more energy from eccentric higher-order modes.

We have shown that without using explicit phasing information, we can still leverage information from the TF morphology using the TF tracks. We developed phenomenological models that would be useful to obtain rapid and coarse constraints on the eccentricity measurement. Such a constraint will help as a first-step indicator to suggest the eccentric nature in the systems which have sufficient inspiral energy and can reduce the computation required to carry out the full-parameter estimation problem. The relevance of this approach improves for future detectors third-generation detectors where we expect long and loud inspiral phase and the eccentric signatures will be even more prominent. Further, the scaling model clearly shows there may not be any degeneracy between higher-order eccentric modes and higher-order modes due to mass asymmetry at least for moderate eccentricities. For a very low-eccentricity system, due to lack of sufficient power in the eccentric higher-order modes, the problem may not be severe.

The method can be further improved if the eccentric waveform model can clearly give the power emitted in each higher-order eccentric mode. For the moment, we have used is EccentricTD model which incorporates eccentricity in an exact manner so the complete harmonic structure is absorbed. Thus, we do not get information about the power in individual eccentric modes. We are investigating on this as a future possibility. The preliminary study with the eccentric, spinning waveform indicates that for low spins and eccentricity, the phenomenological model of ECMM and the scaling factors is robust. At present, we are extending the ECMM to incorporate the aligned spins and eccentric models. 

The TF based ECMM approach shows the direct potential to use nontemplate based approaches, independent of different eccentricity models in the nonspinning or low-spinning cases. In future, with high SNR gravitational wave events, such nonstandard approaches will help to constrain eccentricity in a low-latency approach, in a waveform independent way before going for full Bayesian parameter estimation approach with a waveform model. We have demonstrated the same for nonspinning or low-spinning case here.

\section*{Acknowledgements}
The authors acknowledge Gregorio Carullo, Shubhanshu Tiwari, Srishti Tiwari, Saanika Chowdhury, and Francesco Salemi for important inputs and discussions regarding the manuscript. Thanks to Aswin Suresh and Praveer Tiwari for carrying out model testing for spinning eccentric waveforms and sharing the results. The authors thank the anonymous referee for useful comments. R.H. acknowledges KVPY, DST, India. N.B. acknowledges Inspire division, DST, Government of India. A.P. acknowledges the SERB Matrics Grant No. MTR/2019/001096, SERB-Power-fellowship Grant No. SPF/2021/000036, DST, India and SPARC/2019-2020/P2926/SL, MoE, India for support. This document has LIGO Document No. LIGO-P2300304.

\bibliography{ref}

\end{document}